# Topological protection of multiparticle dissipative transport

Johannes Loehr[1], Michael Loenne[2], Adrian Ernst[3], Daniel de las Heras[3] & Thomas M. Fischer[1]

Topological protection allows robust transport of localized phenomena such as quantum information, solitons and dislocations. The transport can be either dissipative or non-dissipative. Here, we experimentally demonstrate and theoretically explain the topologically protected dissipative motion of colloidal particles above a periodic hexagonal magnetic pattern. By driving the system with periodic modulation loops of an external and spatially homogeneous magnetic field, we achieve total control over the motion of diamagnetic and paramagnetic colloids. We can transport simultaneously and independently each type of colloid along any of the six crystallographic directions of the pattern via adiabatic or deterministic ratchet motion. Both types of motion are topologically protected. As an application, we implement an automatic topologically protected quality control of a chemical reaction between functionalized colloids. Our results are relevant to other systems with the same symmetry.

[1] Experimentalphysik, Institutes of Physics and Mathematics, Universität Bayreuth, Universitätsstraβe 30, Bayreuth 95440, Germany. [2] Mathematik, Institutes of Physics and Mathematics, Universität Bayreuth, Bayreuth 95440, Germany. [3] Theoretische Physik, Institutes of Physics and Mathematics, Universität Bayreuth, Bayreuth 95440, Germany. Correspondence and requests for materials should be addressed to T.M.F. (email: thomas.fischer@uni-bayreuth.de).

Topological invariants are global properties of a system that remain unchanged by local perturbations. A property that depends only on topological invariants is topologically protected and is very robust against local changes. Topological protection is a promising approach to stabilize quantum computing[1] and is used to, for example, maintain robust transport in Hamiltonian systems. Topologically required edge states[2] in a bulk system can support transport of quantum mechanical excitations[3], classical mechanical solitons[4], dislocations[5] and gyroscopic waves[6]. When the edge states are located in a gap of the bulk excitation spectrum, they are protected against scattering into bulk states. Conservation of the Chern number, which is a topological invariant, makes the edge states robust against perturbative interactions. Topological insulators[7], which are based on this concept, conduct at the surface but insulate in bulk. In driven Hamiltonian systems, additional invariants, such as the winding number[8,9] around quasi energy bands, add to the topological variety of possible transport phenomena.

Transport of a collection of classical particles with different properties, such as size, mobility and so on, usually generates a diffuse broadening of the trajectories. Topological protection might be used to transport a broad distribution of particles without dispersion, despite their different properties. High precision multiparticle transport is an important ingredient in, for example, multifunctional lab-on-a-chip devices[10,11].

Topological protection is also possible in driven dissipative (non-Hamiltonian) systems. The interplay between dissipation and topology has been studied in open quantum systems, see for example refs 12,13 for details. In driven dissipative lattices[14–18], transport typically involves the thermal ratchet effect[19–22], that is, biased irreversible jumps between neighbouring potential wells. Complicated correlations between the noise[23], disorder[24] and many particle interactions[25] cause a motion of astonishing complexity. The thermal ratchet mechanism is not robust when transporting simultaneously different types of particles. The complexity makes it hard to maintain control over the transport of one type of particles when adjusting the external drive to control the transport of another particle type.

Here, we show an example of topological protection in a driven dissipative colloidal system. We achieve predictable multiparticle transport of diamagnetic and paramagnetic colloids above a hexagonal magnetic lattice. Using periodic boundary conditions, we describe the unit cell of the lattice as a torus, which defines the action space in which the colloids move. We drive the colloids with periodic modulation loops of an external magnetic field. The direction of the external field defines our control parameter space. The topological correspondence between control and action space is nontrivial, and enables robust, topologically protected, colloidal transport along the lattice vectors. The topological invariant in action space is the set of the two winding numbers around the torus, in close analogy with driven quantum systems[8,9]. We demonstrate experimentally the robustness of the motion and implement a topologically protected quality control of a chemical reaction between functionalized colloids. We also develop a theoretical framework that fully describes the experimental findings. Our results apply to any hexagonal pattern.

## Results

**Colloidal model system.** We use paramagnetic polystyrene core shell and solid polystyrene colloids of average diameters 2.8 and 3.1 µm, respectively, dispersed in a mixture of diluted water-based ferrofluid. The immersion of the colloids renormalizes their effective susceptibilities such that $\chi_{p,eff} > 0$ and $\chi_{d,eff} < 0$ for paramagnetic and diamagnetic colloids, respectively. The colloids immersed in the ferrofluid are placed on top of a magnetically patterned ferrite garnet film (FGF), see Fig. 1a. Spacer beads and a top glass plate create a ferrofluid film of thickness $d = 4.8$ µm. Magnetic boundary conditions at the garnet-ferrofluid and glass-ferrofluid interfaces distort the magnetic field lines (created by the magnetic moments of the colloids) to be parallel to both interfaces. Virtual image dipoles form in the garnet film and the top glass plate and generate a potential that levitates the colloids into the mid-film plane, far away from the FGF, see Fig. 1b. Without the ferrofluid the colloids sediment to the pattern[26,27]. The FGF is characterized by a hexagonal lattice of magnetic bubble domains magnetized normal to the film (saturation magnetization $M_s = 17$ kA m$^{-1}$). The bubbles are immersed in a continuous phase of opposite magnetization. In an external field $H^z_{ext}$ normal to the film, the bubbles grow on the expense of the continuous phase if $H^z_{ext} > 0$ and shrink if $H^z_{ext} < 0$.

**Control space.** We use a homogeneous time-dependent magnetic external field $\mathbf{H}_{ext}(t)$ of constant magnitude, $H_{ext} = 5$ kA m$^{-1}$, to drive the system. Hence, our control space $\mathcal{C}$ is the surface of a sphere. Each point on $\mathcal{C}$ corresponds to a direction of $\mathbf{H}_{ext}$. For reasons that will become clear later, we can divide $\mathcal{C}$ in three regions: the north, the tropics and the south, see Fig. 1c. We call the interface between the tropics and the north (south) as the northern (southern) border. Each border is made of 12 segments. We experimentally perform periodic closed modulation loops $\mathcal{L}_\mathcal{C}$ of the external magnetic field. The period of the modulation is irrelevant provided that it is large enough such that the particles can follow the changes of the potential generated by $\mathbf{H}_{ext}$. There exist loops that induce intercellular colloidal transport. That is, when $\mathcal{L}_\mathcal{C}$ returns to its initial point, the colloids are not in their initial positions but on a different unit cell.

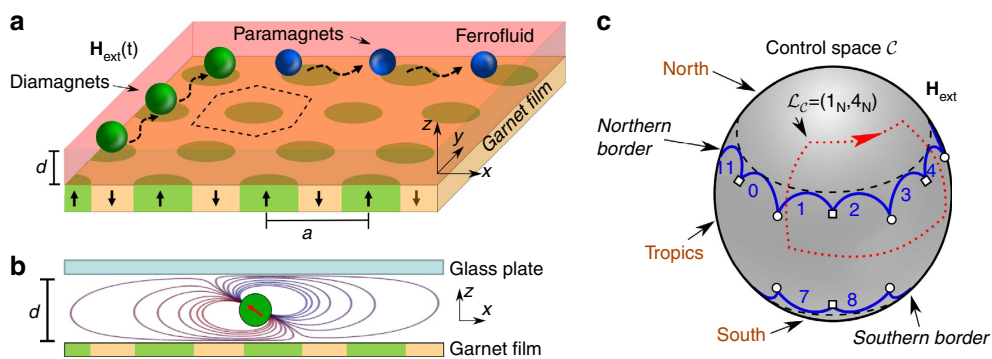

**Figure 1 | Schematic of the system.** (**a**) Hexagonal garnet film with lattice constant $a = 11.6$ µm covered with ferrofluid of thickness $d = 4.8$ µm. One Wigner–Seitz unit cell is marked with a dashed line. By adjusting a closed modulation loop of a spatially homogeneous magnetic field $\mathbf{H}_{ext}(t)$, we have total control over the transport of paramagnetic (blue) and diamagnetic (green) colloids immersed into the ferrofluid. (**b**) Lateral view of the system showing the distortion of the dipolar magnetic field (the field of the garnet pattern is omitted here) of an individual particle immersed in ferrofluid. The field distortion pushes the colloidal particle into the midplane of the ferrofluid film. (**c**) The direction of $\mathbf{H}_{ext}$ varies on the surface of a sphere, defining control space $\mathcal{C}$. Control space can be divided into three regions: the north, the tropics and the south. The northern and southern borders separate the tropics from the north and the south, respectively. Each border consists of 12 segments that we number from 0 to 11. The segments join at special points, indicated by empty circles and squares. $\mathcal{L}_\mathcal{C}$ is an example of a closed modulation loop of $\mathbf{H}_{ext}$ that induces transport of diamagnetic particles along the lattice. The loop crosses the northern border through segments 1 and 4.

**Experimental phase diagram.** Only loops that cross the northern (southern) border of $\mathcal{C}$ induce intercellular transport of the diamagnets (paramagnets). We discuss first the motion of the diamagnets. Let $\mathcal{L}_\mathcal{C} = (i_N, j_N)$ be a loop in $\mathcal{C}$ that crosses the $i$th segment of the northern border from the tropics to the north and returns to the tropics using the $j$th segment, see an example in Fig. 1c. The experimental phase diagram showing the motion of diamagnetic colloids for all possible modulation loops of type $\mathcal{L}_\mathcal{C} = (i_N, j_N)$ is shown in Fig. 2a. The precise shape of the loop is irrelevant, a clear sign of the robustness of the transport. Only the segments of the northern border crossed by $\mathcal{L}_\mathcal{C}$ and their order is important. We can transport the diamagnets along the six fundamental lattice translations plus intracellular transport. Each direction is represented by a different colour in the phase diagram. The clustering of identical colours indicates the topological protection of the transport direction. A rotation of $\mathcal{L}_\mathcal{C}$ by $\pi/3$ around the polar axis, that is, from $\mathcal{L}_\mathcal{C} = (i_N, j_N)$ to $\mathcal{L}_\mathcal{C} = (i_N + 2, j_N + 2)$, is equivalent to rotate the sample by $-\pi/3$, and hence changes the transport direction by $\pi/3$. Therefore, the sixfold symmetry of the pattern guarantees that if transport is possible along one direction then it must also be possible in the other five directions.

There are two types of motion, adiabatic and deterministic ratchet moves. The phase diagram is a checkerboard of alternating adiabatic- and ratchet-squares. In an adiabatic motion, the diamagnets always travel following the minimum generated by the magnetic potential. Hence, the speed of the modulation determines the speed of the colloids along the full trajectory. In contrast, the speed of the modulation loop does not fully determine the speed of the colloids in a ratchet. At some points during the modulation loop, the diamagnets hop between two minima of the magnetic potential at an intrinsic speed that is uncorrelated to the speed of the modulation.

The adiabatic motion is fully reversible. Reversing the modulation from $\mathcal{L}_\mathcal{C} = (i_N, j_N)$ to $\mathcal{L}_\mathcal{C} = (j_N, i_N)$ always reverts the direction of motion, and there is no hysteresis when comparing forward and backward trajectories of the colloids. For example, the loop $\mathcal{L}_\mathcal{C} = (0_N, 4_N)$ transports the diamagnets adiabatically to the left, and the reverse loop $\mathcal{L}_\mathcal{C} = (4_N, 0_N)$ to the right. In a deterministic ratchet motion, reversing the direction of the modulation loop does not usually revert the direction of the transported colloids. $\mathcal{L}_\mathcal{C} = (0_N, 3_N)$, for example, induces a ratchet transporting the diamagnets to the left, but the reverse loop $\mathcal{L}_\mathcal{C} = (3_N, 0_N)$ does not transport the particles to the right. Only some of the modulation loops induce a time reversal ratchet in which reversing the modulation also reverts the direction of motion. See for example, the loops $(0_N, 6_N)$ and $(6_N, 0_N)$ in Fig. 2a. There is always hysteresis in ratchet-like motion between forward and backward trajectories, even in the case of time reversal ratchets.

The dynamics we have discussed for the diamagnets on the northern border holds also for the paramagnets on the southern border of $\mathcal{C}$. The phase diagram of the paramagnets is the same as the one of the diamagnets, cf Fig. 2a, if instead of modulation loops of type $\mathcal{L}_\mathcal{C} = (i_N, j_N)$ we perform modulation loops of type $\mathcal{L}_\mathcal{C} = (i_S, j_S)$. That is, loops that cross the southern border of $\mathcal{C}$ from the tropics to the south using segment $i$ and back to the tropics through segment $j$. An implicit equation to compute the location of the borders is given in the Methods section, and the exact location of the borders is shown in Supplementary Fig. 1.

The northern and southern borders of $\mathcal{C}$ are well separated. Hence, it is easy to transport the diamagnets and paramagnets successively by using a loop $\mathcal{L}_\mathcal{C} = (i_N, j_N, k_S, l_S)$. The loop starts on the tropics and goes to the north of $\mathcal{C}$ crossing the segment $i_N$, then returns to the tropics ($j_N$) and moves to the south ($k_S$). It finally returns to the starting point on the tropics of $\mathcal{C}$ crossing the segment $l_S$. In Fig. 2b, we show polarization microscopy images of the combined transport of six representative modulation loops of the form $\mathcal{L}_\mathcal{C} = (4_N, 0_N, i_S, j_S)$. The loops induce adiabatic transport of diamagnets along the $x$-direction and adiabatic transport of paramagnets along the six possible lattice translations. The trajectories are coloured in pink (cyan) when $\mathcal{L}_\mathcal{C}$ travels on the northern (southern) hemisphere of $\mathcal{C}$.

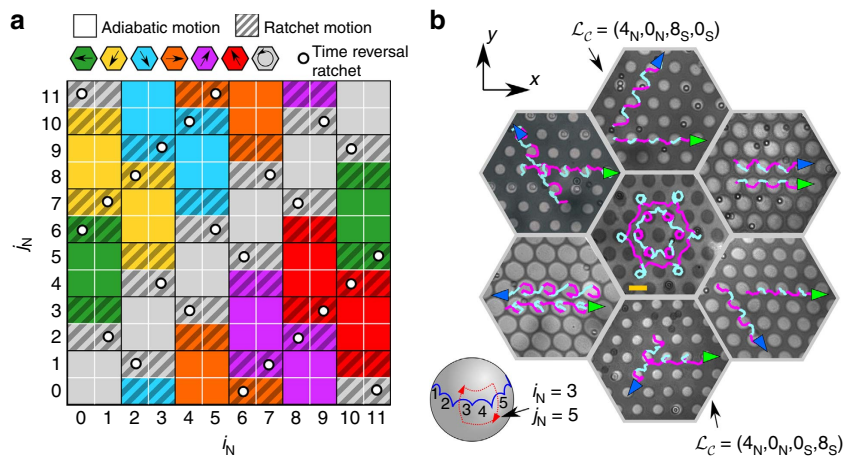

**Figure 2 | Phase diagram and colloidal motion.** (**a**) Experimental phase diagram showing the direction and type of motion of the diamagnets for the fundamental loops $\mathcal{L}_\mathcal{C} = (i_N, j_N)$ crossing the northern border in $\mathcal{C}$. The same diagram holds for the paramagnets if the modulation loops cross the southern border: $\mathcal{L}_\mathcal{C} = (i_S, j_S)$. Each colour corresponds to a direction of motion, as indicated. Non-textured squares indicate adiabatic motion, and striped textured squares indicate ratchet motion. Empty circles mark the time reversal ratchets. (**b**) Polarization microscopy images of the pattern and the diamagnetic and paramagnetic colloidal particles at the end of a transport process. Scale bar (yellow rectangle middle image), 10 μm. The path of one paramagnet (blue arrow) and one diamagnet (green arrow) in $\mathcal{A}$ is depicted in the figure. The pink (cyan) segments of each path indicate the loop in $\mathcal{C}$ is on the southern (northern) hemisphere. The outer images show the transport of diamagnets into the $x$ direction and paramagnets into one of the six crystallographic directions, by using modulation loops of type $\mathcal{L}_\mathcal{C} = (4_N, 0_N, i_S, j_S)$. The middle image is a Franconian folk dance performed by a paramagnetic and diamagnetic couple circulating around a central bubble in opposite sense and with different radius of the hexagon.

We have total control over the colloidal motion, including the ability to programme complex trajectories. An example is given in the centre of Fig. 2b where we use a complex modulation loop such that the paramagnets and diamagnets perform a traditional Franconian folk dance. Videos showing the colloidal motion are provided in Supplementary Movies 1–7.

We next develop the theoretical framework needed to explain the experimental observations we have discussed above. An experimental application will be shown at the end of the Results section.

**Action space.** We call the space accessible to the colloids the action space $\mathcal{A}$. Action space is a two-dimensional periodic hexagonal lattice at a fixed elevation above the FGF. Topologically $\mathcal{A}$ is a torus if we use periodic boundary conditions at the edges of a unit cell of the lattice, see Fig. 3a. Intercellular transport from one unit cell to the next cell via one of the two lattice vectors in real space is the same as one of the two windings around the torus. Loops $\mathcal{L}_\mathcal{A}$ in $\mathcal{A}$ that correspond to intercellular transport of colloids have non-zero winding numbers, and cannot be continuously deformed into a point. That is, lattice translation action loops are non-zero-homotopic. This is not the case in control space. Any modulation loop $\mathcal{L}_\mathcal{C} \subset \mathcal{C}$ can be continuously deformed into any other desired modulation loop. For instance, we can continuously deform $\mathcal{L}_\mathcal{C}$ into a point on $\mathcal{C}$. Therefore, all loops in $\mathcal{C}$ are zero-homotopic.

Here, we have demonstrated that there exist modulation loops $\mathcal{L}_\mathcal{C}$ in control space that induce either adiabatic or deterministic ratchet intercellular transport of the colloids. That is, there are zero-homotopic loops in $\mathcal{C}$ that induce non-zero-homotopic action loops $\mathcal{L}_\mathcal{A}$ with non-vanishing winding number around the torus. To understand how this is possible, we study theoretically the motion of point dipoles in the magnetic potential generated by the garnet and the external field.

**Stationary manifold.** The full dynamics is described by a point $(\mathbf{H}_{ext}, \mathbf{x}_\mathcal{A})$ moving in the product phase space $\mathcal{C} \otimes \mathcal{A}$, where $\mathbf{x}_\mathcal{A} \in \mathcal{A}$ is the position in action space. The energy landscape is given by the magnetic potential $V_m = -\chi_{eff}\mu_0\mathbf{H}^2$, with $\mathbf{H}$ the total magnetic field and $\mu_0$ the vacuum permeability. $\mathbf{H}$ is the sum of the external field $\mathbf{H}_{ext} \in \mathcal{C}$ and the internal field $\mathbf{H}_g(\mathbf{x}_\mathcal{A})$ from the garnet film. The effective susceptibility $\chi_{eff}$ is positive for the paramagnets and negative for the diamagnets. Therefore, the unique scaled-potential $V = H^2$ is enough to qualitatively describe the motion of both types of colloids. The stable points for the diamagnetic (paramagnetic) colloids are the minima (maxima) of $V$. The colloids are far away from the garnet film. Hence, we can approximate the potential by its leading non-constant term at large elevations, which is given by:

$$V \propto \sum_{i=1}^{6} \begin{pmatrix} \mathbf{H}_{ext}^{\|} \\ \tilde{H}_{ext}^z \end{pmatrix} \cdot \begin{pmatrix} \mathbf{q}_{2_i} \sin(\mathbf{q}_{2_i} \cdot \mathbf{x}_\mathcal{A}) \\ q_2 \cos(\mathbf{q}_{2_i} \cdot \mathbf{x}_\mathcal{A}) \end{pmatrix}, \quad (1)$$

where the sum runs only over the six reciprocal lattice vectors of the second Brillouin zone, $\mathbf{q}_{2_i}$, all of which have magnitude $q_2$. The full expression of $V$, at any elevation, is given in the Supplementary Note 1. $\mathbf{H}_{ext}^{\|}$ and $\tilde{H}_{ext}^z = H_{ext}^z/(1+\chi)$ are the components of the external magnetic field in the ferrofluid parallel and normal to the garnet film, respectively. $\chi$ is the magnetic susceptibility of the ferrofluid. $V$ is independent of the details of the FGF, and hence the following theory can be transferred to other systems with the same symmetry. For each value of $\mathbf{H}_{ext}$, the stationary points $(\mathbf{H}_{ext}, \mathbf{x}_\mathcal{A})$ are those for which $\nabla_\mathcal{A} V = 0$, where $\nabla_\mathcal{A}$ indicates the gradient in action space. The set of these points forms the stationary manifold $\mathcal{M}$, which is a two-dimensional manifold in $\mathcal{C} \otimes \mathcal{A}$. Only if $\mathcal{M}$ contains non-zero-homotopic loops, we can achieve intercellular transport. $\mathcal{M}$ can be viewed as the unification of three submanifolds: $\mathcal{M} = \mathcal{M}_+ \cup \mathcal{M}_0 \cup \mathcal{M}_-$. The Hessian matrix is positive definite in $\mathcal{M}_+$ (minima of $V$ and hence stable points for the diamagnets), indefinite in $\mathcal{M}_0$ (unstable saddle points for both colloids) and negative definite in $\mathcal{M}_-$ (maxima of $V$ and hence stable points for the paramagnets). One can show that $\mathcal{M}$ has genus 7 with 3 holes in $\mathcal{M}_0$ and 2 holes in each, $\mathcal{M}_+$ and $\mathcal{M}_-$, see Fig. 3b and Supplementary Fig. 2.

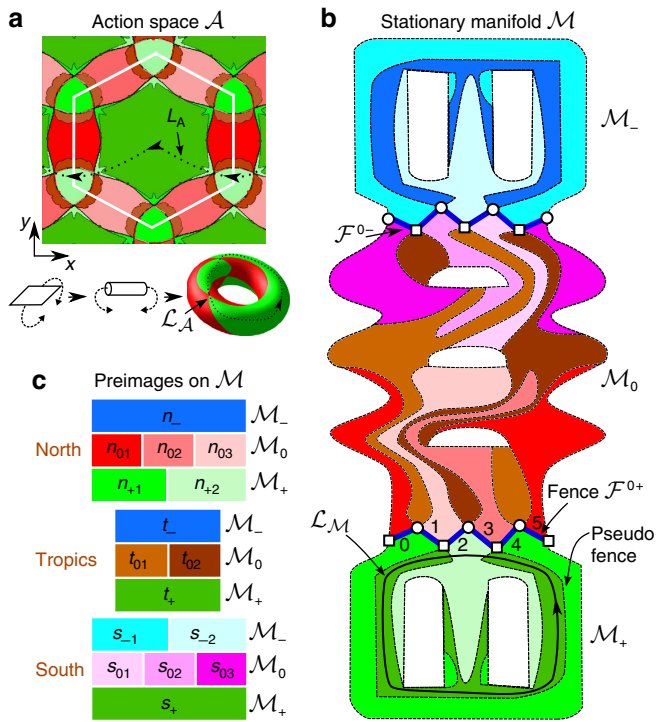

**Figure 3 | Topology.** (a) Action space $\mathcal{A}$ is the space accessible to the colloids, a hexagonal lattice. Using periodic boundary conditions, action space is topologically a torus. (b) Two-dimensional projection of the stationary manifold $\mathcal{M}$, which has genus 7 and it is formed by 16 bijective areas indicated by different colours and listed in c. $\mathcal{M}_+$, $\mathcal{M}_0$ and $\mathcal{M}_-$ are the set of minima, saddle points and maxima of the magnetic potential, respectively. The fence $\mathcal{F}^{0+}$ ($\mathcal{F}^{0-}$) separates $\mathcal{M}_0$ and $\mathcal{M}_+$ ($\mathcal{M}_-$), and it is projected onto the northern (southern) border of control space, cf. Fig. 1c. In **b**, empty squares (circles) on $\mathcal{F}^{0+}$ are triple plus $\mathcal{B}^+$ (zero $\mathcal{B}^0$) bifurcation points, at which 4 bijective areas meet. Three out of these bijective areas lie on $\mathcal{M}_+$ ($\mathcal{M}_0$) in a $\mathcal{B}^+$ ($\mathcal{B}^0$) point. $\mathcal{L}_\mathcal{M}$ (**b**) is an example of a non-zero-homotopic loop that winds around the holes of $\mathcal{M}_+$. The corresponding control loop is $\mathcal{L}_\mathcal{C} = (1_N, 4_N)$. This loop in action space $\mathcal{L}_\mathcal{A}$ induces intercellular transport of the diamagnets along the $-x$ direction, black arrows in **a**. The colours in **a** show the projection of $\mathcal{M}_+$ and $\mathcal{M}_0$ onto action space.

Let $P_\mathcal{C}$ be the projection that maps any point in $\mathcal{C} \otimes \mathcal{A}$ into control space. A key point is that $P_\mathcal{C}$ is multifold on $\mathcal{M}$, that is, several points $\mathbf{x}_\mathcal{M} = (\mathbf{H}_{ext}, \mathbf{x}_\mathcal{A})$ are mapped on the same point $\mathbf{H}_{ext} \in \mathcal{C}$. The north, the south and the tropics of $\mathcal{C}$, cf. Fig. 1c, have different multiplicity of preimages on $\mathcal{M}$. The multiplicity changes at the borders of $\mathcal{C}$ via generation or annihilation of pairs involving one saddle point and one minimum or one maximum. We can divide $\mathcal{M}_{0,\pm}$ into a collection of bijective areas, $\{n_{+1}, n_{+2}, t_+, s_+\} \subset \mathcal{M}_+$, $\{n_{01}, n_{02}, n_{03}, t_{01}, t_{02}, s_{01}, s_{02}, s_{03}\} \subset \mathcal{M}_0$ and $\{n_-, t_-, s_{-1}, s_{-2}\} \subset \mathcal{M}_-$. Each area has exactly one preimage of either the north, the tropics or the south of $\mathcal{C}$. The letter indicates if the area is projected onto the north (n), the tropics (t) or the south (s) of $\mathcal{C}$. These areas are listed in Fig. 3c (with the colours corresponding to the colouring of $\mathcal{M}$, Fig. 3b). The first subindex (0, +, −) indicates if the area lies on $\mathcal{M}_0$, $\mathcal{M}_+$ or on $\mathcal{M}_-$. The second subindex labels the areas in case more than one area share the same letter and first subindex.

**Fences and bifurcation points.** We call the boundary between $\mathcal{M}_0$ and $\mathcal{M}_+$ ($\mathcal{M}_-$) as the northern $\mathcal{F}^{0+}$ (southern $\mathcal{F}^{0-}$) fence, see Fig. 3b. A segment of $\mathcal{F}^{0+}$ separates a northern area on $\mathcal{M}_+$ ($n_{+,1}$ or $n_{+,2}$) from a northern area on $\mathcal{M}_0$ ($n_{0,1}, n_{0,2}$ or $n_{0,3}$) and starts and ends at vertices that are bifurcation points. Four different bijective areas in $\mathcal{M}$ meet at a bifurcation point, see Fig. 3b. There are three types of bifurcation points: triple zero bifurcation points $\mathcal{B}^0$, where three out of the four areas meeting at the point are on $\mathcal{M}_0$, and triple plus $\mathcal{B}^+$ (minus $\mathcal{B}^-$) bifurcation points, where three out of the four areas meeting at the point are on $\mathcal{M}_+$ ($\mathcal{M}_-$). In total, each fence has 12 bifurcation points that alternate between $\mathcal{B}^0$ and $\mathcal{B}^+$ or $\mathcal{B}^-$, depending on the fence. No further points where more than two areas meet on $\mathcal{M}$ exist. The vertices on the fence are the only bifurcation points on $\mathcal{M}$. The projection $P_\mathcal{C}$ maps each of the 12 segments of $\mathcal{F}^{0+}$ ($\mathcal{F}^{0-}$) onto one segment of the northern (southern) border of $\mathcal{C}$, see Fig. 1c. $P_\mathcal{C}$ also maps the bifurcation points on $\mathcal{F}^{0+}$ ($\mathcal{F}^{0-}$) onto 12 points at the northern (southern) border of $\mathcal{C}$ where two segments join. As $P_\mathcal{C}$ is multifold on $\mathcal{M}$, the preimage of the borders of $\mathcal{C}$ are the fences and other lines that we call the pseudo fences. The preimage of the projection of the bifurcation points are the bifurcation points and other points that we call pseudo bifurcation points. The pseudo fences separate different bijective areas on $\mathcal{M}$, and are also divided in 12 segments, which are separated by pseudo bifurcation points. We label the segments of the borders of $\mathcal{C}$, and the segments of the fences and pseudo fences in $\mathcal{M}$ from 0 to 11. A segment $i$ on $\mathcal{M}$ is projected onto the segment $i$ on $\mathcal{C}$. Therefore, if we cross the $i$th segment of the border in $\mathcal{C}$, we cross several $i$th segments of fences and pseudo fences on $\mathcal{M}$.

**Adiabatic motion.** We next explain the adiabatic transport of diamagnets, similar arguments apply for the paramagnets. To achieve adiabatic transport of diamagnets, we need a modulation loop $\mathcal{L}_\mathcal{C}$ with a preimage loop $\mathcal{L}_\mathcal{M}$ in $\mathcal{M}$ lying entirely in $\mathcal{M}_+$, such that the diamagnets can adiabatically follow the minimum of the magnetic potential. In addition, $\mathcal{L}_\mathcal{M}$ has to be non-zero-homotopic, that is, it has to wind around at least one of the two holes in $\mathcal{M}_+$. This non-zero-homotopic loop is then projected onto a loop in $\mathcal{A}$ that can be non-zero-homotopic, and induce intercellular transport. As we have already shown adiabatic motion along any lattice direction, $\mathbf{a} = w_1\mathbf{a}_1 + w_2\mathbf{a}_2$, with $\mathbf{a}_i$ the basic lattice vectors in $\mathcal{A}$, is possible. Each transport direction corresponds to a value of the set of the two winding numbers $\{w_1, w_2\}$ around the hole in $\mathcal{A}$. Hence, our topological invariant is the set of winding numbers in $\mathcal{A}$. In $\mathcal{M}$ there are 7 holes, and hence 14 winding numbers. The sum of any winding number in $\mathcal{M}$ over all loops $\mathcal{L}_\mathcal{M}$ corresponding to a given loop in $\mathcal{C}$ is zero since all loops in $\mathcal{C}$ are zero-homotopic. We can only achieve a non-zero-homotopic loop in $\mathcal{M}$ by first joining two zero-homotopic loops in $\mathcal{M}$, and next disjoining them into two loops with opposite winding numbers. The detailed explanation is shown next.

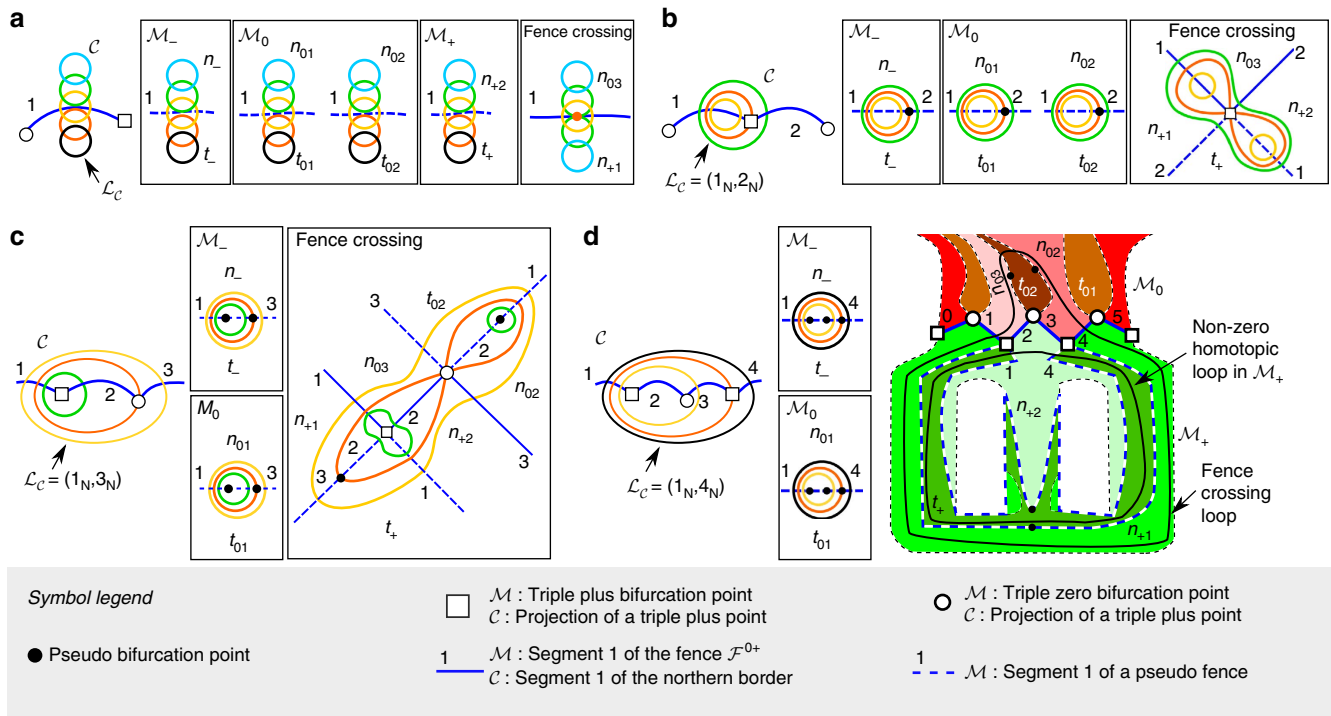

**Figure 4 | Joining and disjoining loops in $\mathcal{M}$.** Schematic of different modulation loops $\mathcal{L}_\mathcal{C}$ in control space $\mathcal{C}$ and their corresponding preimage loops on the stationary surface $\mathcal{M}$. (**a**) $\mathcal{L}_\mathcal{C}$ crosses the first segment of the northern border of $\mathcal{C}$. When $\mathcal{L}_\mathcal{C}$ touches the border (red loop) a pair of a minimum and a saddle point is created in $\mathcal{M}$ (red point). When $\mathcal{L}_\mathcal{C}$ crosses the border twice (yellow loop), a loop crossing the fence in $\mathcal{M}$ (fence-crossing loop) is created (yellow loop). This fence-crossing loop lies in both $\mathcal{M}_+$ and $\mathcal{M}_0$. (**b**) We enlarge $\mathcal{L}_\mathcal{C}$ such that it encircles the projection of a triple plus bifurcation point. In $\mathcal{M}$, the fence-crossing loop joins with the loop in $\mathcal{M}_+$. No loop lies entirely in $\mathcal{M}_+$. (**c**) $\mathcal{L}_\mathcal{C}$ encircles the projection of two bifurcation points, one $\mathcal{B}^+$ and one $\mathcal{B}^0$. The fence-crossing loop joins again with another loop that this time lies in $\mathcal{M}_0$. (**d**) $\mathcal{L}_\mathcal{C}$ encircles now the projection of two $\mathcal{B}^+$ and one $\mathcal{B}^0$ bifurcation points. The four areas meeting at the second $\mathcal{B}^+$ point ($n_{+1}, n_{+2}, t_+$ and $n_{02}$) were already joined in the fence-crossing loop. As a result, the fence-crossing loop disjoins into two loops, that in this case are non-zero homotopic with opposite winding numbers. One of the disjoint loops lies in $\mathcal{M}_+$ and induces intercellular adiabatic motion. All loops in **a**–**c** are zero-homotopic.

Consider the preimage in $\mathcal{M}$ of the modulation loop $\mathcal{L}_\mathcal{C} = (1_N, 1_N)$. We show a schematic of $\mathcal{L}_\mathcal{C}$ and all its preimage loops in $\mathcal{M}$ in Fig. 4a. If $\mathcal{L}_\mathcal{C}$ is entirely in the tropics of $\mathcal{C}$ (black loop) there are four zero-homotopic preimage loops on $\mathcal{M}$. One is in $\mathcal{M}_-$, two in $\mathcal{M}_0$ and another one in $\mathcal{M}_+$. When $\mathcal{L}_\mathcal{C}$ touches the northern border of $\mathcal{C}$ (red loop), a pair of a minimum and a saddle point is generated at the fence $\mathcal{F}^{0+}$. As $\mathcal{L}_\mathcal{C}$ crosses the northern border of $\mathcal{C}$ (yellow loop), the minimum-saddle point pair deforms into a fifth (zero-homotopic) loop on $\mathcal{M}$ that crosses the fence $\mathcal{F}^{0+}$. This new loop eventually disjoins into two new loops, one on $\mathcal{M}_0$ and one on $\mathcal{M}_+$, when $\mathcal{L}_\mathcal{C}$ fully enters the north of $\mathcal{C}$ (blue loop). At each stage in $\mathcal{C}$, the other four loops on $\mathcal{M}$ smoothly pass through different pseudo fences on $\mathcal{M}$. All loops on $\mathcal{M}$ produced with modulation loops $\mathcal{L}_\mathcal{C} = (i_N, i_N)$ are zero-homotopic and therefore do not produce transport in $\mathcal{A}$. The specific bijective areas covered by the loops on $\mathcal{M}$ depend on the segment of the border that we cross in $\mathcal{C}$. A figure showing the bijective areas that meet at each segment of fences and pseudo fences is given in Supplementary Fig. 3.

Let us now deform $\mathcal{L}_\mathcal{C}$ such that it finally encircles the projection of a triple plus bifurcation point, see Fig. 4b. The final loop is $\mathcal{L}_\mathcal{C} = (1_N, 2_N)$. When $\mathcal{L}_\mathcal{C}$ crosses the projection of $\mathcal{B}^+$, the corresponding loop $\mathcal{L}_\mathcal{M}$ crossing the fence on $\mathcal{M}$ joins with the pseudo fence-crossing loop on $\mathcal{M}_+$. The result is a new loop that crosses the fence and passes through four areas on $\mathcal{M}$. This loop lies in both $\mathcal{M}_0$ and $\mathcal{M}_+$. As no other loop entirely lies on $\mathcal{M}_+$, the diamagnets will follow a ratchet motion, leaving the stationary surface $\mathcal{M}$ when the loop crosses the fence towards $\mathcal{M}_0$. We will explain the ratchet motion later on. The winding number of the joint fence-crossing loop on $\mathcal{M}$ is the sum of the winding numbers of the loops before the joining. In this case, the joining loops are zero-homotopic and hence the joint loop is also zero-homotopic and induces no transport in $\mathcal{A}$.

In Fig. 4c, we further expand the modulation loop such that it encircles the following projection of a bifurcation point, a $\mathcal{B}^0$. The final loop is $\mathcal{L}_\mathcal{C} = (1_N, 3_N)$. In $\mathcal{M}$, we again join the fence-crossing loop with a pseudo fence-crossing loop that now lies in $\mathcal{M}_0$. The result is, as in the previous case, a zero-homotopic fence-crossing loop.

We continue expanding the modulation loop such that it finally encircles the projection of two $\mathcal{B}^+$ points with $\mathcal{L}_\mathcal{C} = (1_N, 4_N)$, see Fig. 4d. Now, all four areas that meet at the second $\mathcal{B}^+$ bifurcation point in $\mathcal{M}$ are already joined in the fence-crossing loop. Therefore, crossing this bifurcation point disjoins the fence-crossing loops in two loops. The disjoint loops are no longer zero-homotopic. They have winding numbers with equal magnitude but opposite sign such that the sum is zero. One of the disjoint loops lies entirely in $\mathcal{M}_+$, crosses the segments 1 and 4 of the pseudo fence between $n_{+2}$ and $t_+$ and winds around the holes in $\mathcal{M}_+$. This loop is projected into a non-zero-homotopic loop in $\mathcal{A}$ that induces adiabatic transport of the diamagnets along the $-x$ direction.

Encircling the next projection of a $\mathcal{B}^+$ point, $\mathcal{L}_\mathcal{C} = (1_N, 6_N)$, joins again the loop in $\mathcal{M}_+$ with a fence-crossing loop and creates a ratchet motion. The adiabatic transport is recovered by encircling a further projection of a $\mathcal{B}^+$ with $\mathcal{L}_\mathcal{C} = (1_N, 8_N)$. This disjoins the fence-crossing loop and generates a new non-zero-homotopic loop in $\mathcal{M}_+$. This new loop crosses segments and pseudo fences in $\mathcal{M}_+$ that are different than the previous non-zero-homotopic loop, and induces transport in a different lattice direction.

**Deterministic ratchet motion.** We next explain why the deterministic ratchet is topologically protected and its fundamental role in the phase diagram. A ratchet motion occurs if there is no loop that lies entirely on $\mathcal{M}_+$. In this case, the minimum of the magnetic potential that transports the diamagnets disappears at the fence, and the particles leave the stationary manifold $\mathcal{M}$ jumping to another minimum.

Our modulation is adiabatic, that is, the relaxation time of the colloids in the cage around the minimum is orders of magnitude faster than the period of the modulation. Hence, if the diamagnets are on $\mathcal{M}_+$, they follow the minimum of the potential with a dynamics given by the modulation. If, on the contrary, the diamagnets are not on $\mathcal{M}_+$, they move along the path of steepest descend of an effectively frozen potential in $\mathcal{C}$. This path brings the diamagnets back to $\mathcal{M}_+$.

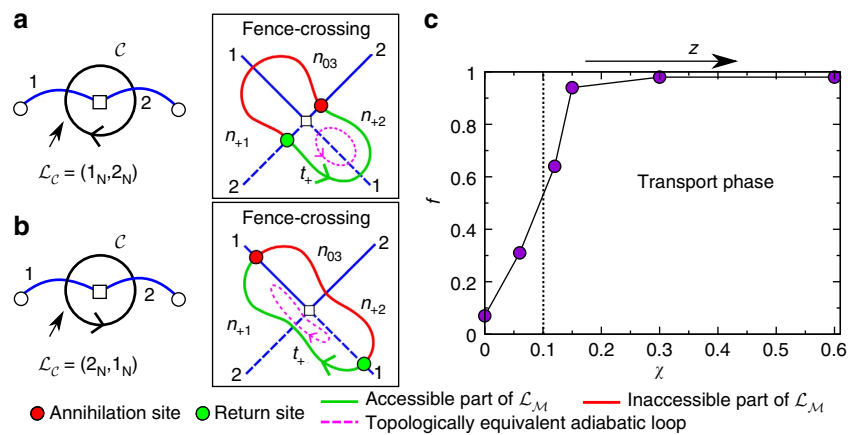

**Figure 5 | Deterministic ratchet motion and elevation of the colloids.** (**a**) A loop in $\mathcal{C}$ encircling the projection of a triple plus bifurcation point and the corresponding fence-crossing loop in $\mathcal{M}$. (**b**) Same as in **a** but for a modulation loop in the opposite direction. The arrows in **a**,**b** indicate the starting points and the directions of the loops in $\mathcal{C}$ and $\mathcal{M}$. The violet dashed loop is an adiabatic loop topologically equivalent to the deterministic ratchet loop. It is formed by gluing the annihilation and return sites of the ratchet loop. (**c**) Fraction of transported colloids $f$ as a function of the magnetic susceptibility $\chi$ of the ferrofluid. $f$ is computed by counting how many colloids out of 100 have been successfully transported after a modulation loop. The vertical dotted line approximately marks the transition between transport and non-transport phases.

Consider again the modulation loop $\mathcal{L}_\mathcal{C} = (1_N, 2_N)$ that encircles the projection of a $\mathcal{B}^+$ point and creates a ratchet. In Fig. 5a, we plot the loop in $\mathcal{C}$ and the corresponding fence-crossing loop in $\mathcal{M}$. We start $\mathcal{L}_\mathcal{C}$ in the tropics of $\mathcal{C}$. In $\mathcal{M}$, the diamagnets follow the segment in $t_+$ of the loop. Next, $\mathcal{L}_\mathcal{C}$ crosses the first segment of the northern border, and the diamagnets cross the first segment of the pseudo fence between $t_+$ and $n_{+2}$ in $\mathcal{M}$. Finally, $\mathcal{L}_\mathcal{C}$ crosses the second segment of the border. At this point, the loop in $\mathcal{M}$ touches the fence. The minimum in $n_{+2}$ that adiabatically transported the particles annihilates with a saddle point and disappears. The colloids leave the stationary surface $\mathcal{M}$ at the annihilation site. Diamagnets follow now the path of steepest descend and are brought back to $\mathcal{M}_+$ through the return site, see Fig. 5a. Hence, the fence-crossing loop in $\mathcal{M}$ can be divided into accessible and inaccessible parts. The particles can stay only in the accessible part. The path of steepest descend connecting the annihilation and the return sites is topologically trivial. It cannot change the homotopy class of the adiabatic loop that emerges by taking the accessible part of the loop and gluing both ends, annihilation and return sites, together, see Fig. 5a. The reason is that the path of steepest descend develops in a continuous manner from the bifurcation point $\mathcal{B}^+$. To understand this, imagine we make $\mathcal{L}_\mathcal{C}$ smaller and smaller but always encircling the projection of $\mathcal{B}^+$. Then, annihilation and return sites come closer and closer to each other, and eventually meet at the bifurcation point. This argument holds for any other ratchet motion in the system. A ratchet loop is always topologically protected by an adiabatic loop. Both loops have the same homotopy class, and therefore the same direction of motion.

Let us now revert the direction of $\mathcal{L}_\mathcal{C}$, see Fig. 5b. The accessible part of the loop and the annihilation and return sites change. The forward and backward adiabatic loops that protect the ratchets are different, but induce transport in opposite directions. Therefore, the ratchet is time reversal. Reverting the modulation reverts the colloidal motion. There is, however, hysteresis since forward and backward loops differ in the path of steepest descend, and in the segments being crossed in $\mathcal{M}$. Usually, forward and backward loops are protected by adiabatic loops that induce transport in different, non-opposite, directions, resulting in a non-time reversal ratchet.

Ratchets play a fundamental role in the system. The homotopy class of an adiabatic loop, which lies on $\mathcal{M}_+$, cannot be changed by continuous deformations. Therefore, the direction of transport cannot change if the motion remains adiabatic (note that all neighbouring adiabatic loops in the phase diagram of Fig. 2a induce transport in the same direction). It is only via ratchets that we can change the homotopy class of a loop and hence the transport direction. See, for example, in Fig. 2a, the ratchet loop $\mathcal{L}_\mathcal{C} = (1_N, 2_N)$ (protected by the adiabatic loop $(1_N, 1_N)$) and the ratchet loop $(1_N, 3_N)$ (protected by $(1_N, 4_N)$). The topological transition that changes the transport direction occurs when $\mathcal{L}_\mathcal{C}$ encircles the projection of a $\mathcal{B}^0$ point (Supplementary Fig. 4).

Theory and experiments are in perfect agreement. The above theory predicts exactly the same phase diagram we have found experimentally, cf Fig. 2a. In addition, we have also performed Brownian dynamic simulations of paramagnetic and diamagnetic particles moving in the potential given by equation (1). The simulations are also in perfect agreement with the theory and the experiments. The simulation allows us to introduce thermal noise in the system. We have verified that the topological protection is very robust against thermal fluctuations. When the noise is very high, such that it erases the energy landscape, the topological protection is lost. The degradation of the topological protection starts at both interfaces between different types of motion in the phase diagram: adiabatic-ratchet interface and the interface between deterministic ratchets along different directions.

The transport direction is also robust against other perturbations, such as the precise shape and speed of the modulation loop, changes in size, mobility and magnetic susceptibility of the colloids, and changes in the pattern that do not affect its symmetry (for example, the shape of the bubbles). Most strikingly, the directions of the ratchets are protected, that is, the topology of the stationary surface determines not only the direction of the adiabatic motion but also the non-equilibrium ratchet motion.

There are always operations that break the topological protection. In our system, we can break the protection by changing the topology of $\mathcal{M}$ as we describe next.

**Elevation above the garnet**. We return now to the experiments. We have discussed the transport of colloids at elevations $z$ far away from the garnet film so that the potential is given by equation (1). At low elevations, the field created by the garnet is very strong compared with the external magnetic field and the potential is given by that of the pattern alone. In this situation, the different parts of $\mathcal{M}$ are disconnected manifolds and have a trivial topology (spheres) missing the requirements for topological transport.

Depending on the dilution of the ferrofluid, the image-dipole potential may or may not overcome the gravitational potential. Hence, controlling the ferrofluid susceptibility gives direct control over the colloidal elevation $z$ above the garnet film. Consider a loop in $\mathcal{C}$ that induces lattice translations if the colloids are at high elevations. By performing the same loop for different ferrofluid-water compositions, and hence varying the susceptibility $\chi$, we can observe the transition from non-zero-homotopic loops in $\mathcal{A}$ to zero-homotopic loops. The results are shown in Fig. 5c. The topological transition takes place at $\chi \approx 0.1$. For $\chi \lesssim 0.1$, the particles descend below a critical elevation $z_c$, and the transition to the non-transport phase occurs. Above $z_c$ the effects of the hexagonal pattern are the same for any $z$, and topologically protected modulation loops work for any hexagonal pattern, independently of the fine details. By decreasing the elevation below $z_c$, we remove the holes of $\mathcal{M}$ inducing a topological transition. This plays the role of gap closure in the dispersion relation of wave-like systems[28].

**Application**. We use the topological protection to implement an experimental internal quality control of a chemical reaction. We consider the hybridization reaction between two complementary single-stranded oligo nucleotides of DNA, which we attach to the paramagnetic and diamagnetic colloids. If the hybridization is successful, the paramagnet ($p$) and diamagnet ($d$) irreversibly bind to form a quadrupole ($q$)

$$p + d \rightarrow q. \qquad (2)$$

We want to emulate the conditional command: if the reaction is successful, then transport the product $q$ along a given direction $a_q$, otherwise transport the educts $p$ and $d$ along directions $a_p$ and $a_d$, respectively.

We have already shown how to induce topologically protected transport of the educts (dipoles). The product of the reaction is a quadrupole that senses the quadrupolar potential $V_q = -(\nabla_A V)^2$. The modulation loops $\mathcal{T}_p \subset \mathcal{C}$ and $\mathcal{T}_d \subset \mathcal{C}$ for the transport of the educts can be chosen such that they do not affect the motion of the quadrupoles. We also find an appropriate modulation loop $\mathcal{T}_q \subset \mathcal{C}$ that transports the quadrupoles in the required direction without affecting the dipoles.

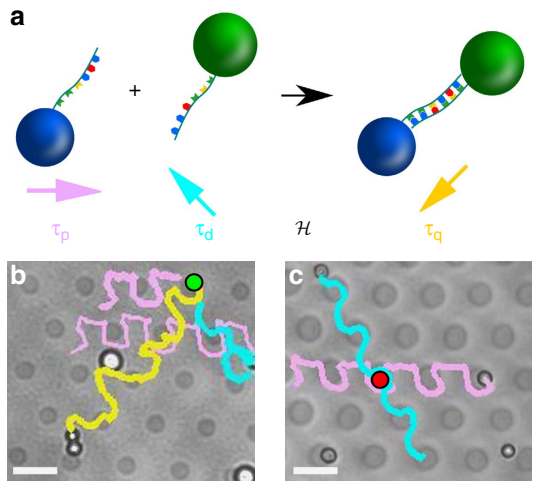

**Figure 6 | Application.** (**a**) Schematic of the hybridization reaction and the emulation of a conditional command that transport each type of particle in different directions. Polarization images of a successful and non-successful reaction are shown in **b**,**c**, respectively. The trajectories of the diamagnets, paramagnets and quadrupoles are highlighted in pink, cyan and yellow, respectively. The green (red) circle indicates the area where the dipoles meet and the hybridization takes place (fails). Scale bars (**b**,**c**), 10 µm.

The paramagnets and diamagnets reside on opposite parts ($\mathcal{M}_-$ and $\mathcal{M}_+$) of the stationary surface and, in the presence of the pattern, do not approach each other in $\mathcal{A}$ to allow the hybridization. To bring the colloids together, we perform a non-adiabatic hybridization loop $\mathcal{H} \subset \mathcal{C}$ around the equator of $\mathcal{C}$ with a very high angular frequency. Hence, the colloids cannot follow the instantaneous potential and feel an almost flat averaged potential. Effectively, we erase the potential of the pattern such that the weak dipolar attractive interaction is enough to bring paramagnets and diamagnets together. The colloids meet in a bubble and rotate around each other such that hybridization is possible. After hybridization, the bond is strong enough to resist the magnetic stress exerted by the potential of the next modulation loops such that the bond is irreversible. The particle pair remains inseparable and behaves like a stable quadrupole. The entire modulation is a loop of the type $\mathcal{T}_p \mathcal{T}_d \mathcal{H} \mathcal{T}_q$, see Fig. 6a. We show the transport of the colloids after a successful and a non-successful hybridization in Fig. 6b,c, respectively. Videos are provided in Supplementary Movies 8–9. This quality control is internal as we do not change the modulation loop after we have measured externally whether the reaction was successful. The quality control works without active intervention.

## Discussion

When the modulation loop is in the north of $\mathcal{C}$, the magnetic potential presents two minima, one in $n_{+1}$ and one in $n_{+2}$ that are projected onto different points on $\mathcal{A}$. No colloidal transport between the minima exists as the potential barrier is too high. Phase space is hence divided into different nonergodic regions, and thermal equilibration only occurs over the cage around each minimum. The cage can only be left in a ratchet-like motion when the modulation loop touches the fence. Hence, our ratchet is associated with an ergodic-nonergodic transition, and might serve as a model for the cage effect in supercooled liquids[29] and glasses[30].

Dissipation has been used in open quantum systems[12] to isolate topologically protected edge modes from a bath. The final state of the edge modes is nevertheless dissipation free.

In contrast, our system must be driven to maintain the transport. Moreover, the ratchet effect (crucial to change the direction of motion) is intrinsically dissipative and cannot, in general, be described with an effective Hamiltonian.

In one-dimensional systems, one can prove that thermal ratchets[31], in which the potential evolves in time adiabatically, are always time reversal ratchets. We have shown here an example in two dimensions of a non-time reversal ratchet in an adiabatically evolving potential.

The construction of the stationary surface $\mathcal{M}$ and the mappings to action and control space is completely general and can be used for any potential. Other potentials might or might not support topologically protected transport, depending on the topological properties of $\mathcal{M}$. Our results are directly transferable to any system with hexagonal symmetry, and a potential proportional to the square of a field, which satisfies the Laplace equation.

High-quality magnetic bubbles lattices, like the one we have used here, have been studied extensively[32] and hence the technology for its fabrication is well known. In addition, we note that any patterned substrates, such as lithographic magnetic patterns[33], will induce similar transport.

## Methods

**Experimental preparation and measurements.** The FGF films were grown by Tom Johansen (Oslo) via liquid phase epitaxy. We use the water-based ferrofluid EMG 707 from FerroTec GmbH, Germany. We dilute the ferrofluid with water. The final magnetic susceptibility is $\chi \approx 0.6$. The time-dependent magnetic field is generated by three coils, following the ideas presented in ref. 34. Each coil controls the magnetic field along one of the three Cartesian axis. The current through the coils is provided by three phase-locked channels of programmable waveform generators (TTi TG1244) via three bipolar (KEPCO 20–50GL) amplifiers. The system is monitored via polarization microscopy. The pattern is visualized via the polar Faraday effect, and the colloids via ordinary transmission microscopy. Modulation loops in control space are programmed on a computer and transferred to the waveform generators.

For the hybridization reaction, we functionalize the colloids with streptavidin. The diamagnets and paramagnets are immersed separately into two solutions of biotinylated and complementary oligonucleotides. The complementary sequences are 5′-/5Bio/TCACTCAGTACGATATGCGGCACAG-3′ and 5′-/5Bio/CTGTGC CGCATATCGTACTGAGTGA-3′.

**Topology of the stationary manifold.** We first find the projection of the bifurcation points and the fences onto action space. Next, we map action space into control space, so that we obtain the projection of the fences and bifurcation points in $\mathcal{C}$. With these projections we compute the vertices $v = 96$, edges $e = 124$ and areas $a = 16$ of $\mathcal{M}$. The Euler characteristic of $\mathcal{M}$ is $\chi(\mathcal{M}) = v - e + a = -12$, and it has genus $g(\mathcal{M}) = 1 - \chi(\mathcal{M})/2 = 7$. The mapping of $\mathcal{A}$ into $\mathcal{C}$ also allows the determination of how the bijective areas are glued in $\mathcal{M}$. Further details are given below.

**Projection of the fence.** We use coordinates

$$\mathbf{x}_A = (x_1 \mathbf{a}_1, x_2 \mathbf{a}_2), \quad (3)$$

in action space $\mathcal{A}$, where $\mathbf{a}_1$ and $\mathbf{a}_2$ are the basic lattice vectors of the hexagonal lattice. In control space, we use coordinates

$$\mathbf{H}_{\text{ext}} = H_{\text{ext}}(\cos\phi \sin\theta, \sin\phi \sin\theta, \cos\theta), \quad (4)$$

where the azimuthal angle $\phi$ is measured with respect to the direction of $\mathbf{a}_1$ and the polar angle $\theta$ with respect to the $z$-direction. Consider the unit vectors $\hat{\mathbf{e}}_i(x_1, x_2) = \partial_i \mathbf{H}_g / |\partial_i \mathbf{H}_g|$, $i = 1, 2$, where $\mathbf{H}_g$ is the magnetic field of the garnet film. As we have seen, the leading term of the magnetic potential is $V \propto \mathbf{H}_{\text{ext}} \cdot \mathbf{H}_g$ and the stationary points are those for which $\nabla_A V = 0$. Then, a point $(\mathbf{H}_{\text{ext}}, \mathbf{x}_A)$ in $\mathcal{C} \otimes \mathcal{A}$ is stationary, and hence lies on $\mathcal{M}$, if the direction of $\mathbf{H}_{\text{ext}}$ is perpendicular to both $\hat{\mathbf{e}}_1$ and $\hat{\mathbf{e}}_2$. Therefore, a point in $\mathcal{A}$ with coordinates $(x_1, x_2)$ in the basis $(\mathbf{a}_1, \mathbf{a}_2)$ has two stationary preimages in $\mathcal{M}$ that correspond to external magnetic fields $\mathbf{H}_{\text{ext}}^{(s)}(x_1, x_2) = \pm H_{\text{ext}}(\hat{\mathbf{e}}_1 \times \hat{\mathbf{e}}_2)/|\hat{\mathbf{e}}_1 \times \hat{\mathbf{e}}_2|$. The superscript $(s)$ in $\mathbf{H}_{\text{ext}}^{(s)}(x_1, x_2)$ indicates that this field makes the point $(x_1, x_2)$ in $\mathcal{A}$ stationary. Consider now the Hessian matrix,

$$\nabla_A \nabla_A V = \begin{pmatrix} \mathbf{H}_{\text{ext}}^{(s)} \cdot \partial_1 \partial_1 \mathbf{H}_g & \mathbf{H}_{\text{ext}}^{(s)} \cdot \partial_1 \partial_2 \mathbf{H}_g \\ \mathbf{H}_{\text{ext}}^{(s)} \cdot \partial_2 \partial_1 \mathbf{H}_g & \mathbf{H}_{\text{ext}}^{(s)} \cdot \partial_2 \partial_2 \mathbf{H}_g \end{pmatrix}. \quad (5)$$

When crossing the fence on $\mathcal{M}$ from $\mathcal{M}_0$ to $\mathcal{M}_+$, a saddle point changes to a minimum. Hence, the determinant of the above Hessian matrix must vanish at the fence, $\|\nabla_\mathcal{A}\nabla_\mathcal{A} V\|_\text{fence}=0$. In this way, we find an implicit equation for the projection of the fences in action space. Both fences, $\mathcal{F}^{0+}$ and $\mathcal{F}^{0-}$, are projected into the same region in $\mathcal{A}$ with coordinates $(x_{1,\text{f}}, x_{2,\text{f}})$ given implicitly by:

$$0 = F^2 - 2SF - 3S^2 - 4(f_1 f_2 + f_1 f_3 + f_2 f_3), \quad (6)$$

where

$$\begin{aligned}
F &= f_1 + f_2 + f_3,\\
S &= c_1 + c_2 + c_3,\\
f_i &= 1 + c_i + c_i^2, \; i = 1, 2, 3,\\
c_1 &= \cos(2\pi x_{1,\text{f}}),\\
c_2 &= \cos(2\pi x_{2,\text{f}}),\\
c_3 &= \cos(2\pi [x_{1,\text{f}} + x_{2,\text{f}}]).
\end{aligned} \quad (7)$$

The projection of the fence $\mathcal{F}^{0+}$ in control space, that is, the northern border on $\mathcal{C}$ is given by:

$$\begin{aligned}
\theta &= \operatorname{atan}\left(\frac{\sqrt{H_1^2 + 2H_1 H_2 \cos(\pi/3) + H_2^2}}{H_0 \sin(\pi/3)}\right),\\
\phi &= \operatorname{atan}\left(\frac{\sin(\pi/3)}{H_1/H_2 + \cos(\pi/3)}\right),
\end{aligned} \quad (8)$$

where

$$\begin{aligned}
H_1 &= c_3(s_1 - s_2) + c_2(s_1 + s_3),\\
H_2 &= c_3(s_2 - s_1) + c_1(s_2 + s_3),\\
H_0 &= c_1 c_2 + c_2 c_3 + c_3 c_1,\\
s_1 &= \sin(2\pi x_{1,\text{f}}),\\
s_2 &= \sin(2\pi x_{2,\text{f}}),\\
s_3 &= \sin(2\pi [x_{1,\text{f}} + x_{2,\text{f}}]).
\end{aligned} \quad (9)$$

The coordinates of the southern border on $\mathcal{C}$ are then obtained via the transformation $\theta \to \pi - \theta$ and $\phi \to \phi - \pi$. Supplementary Fig. 1 shows the projection of the fences in the $\phi - \theta$ plane of control space.

**Projection of bifurcation points.** Four bijective areas meet at a bifurcation point in $\mathcal{M}$. Four segments (two fence segments and two pseudo-fence segments) form the branches that bifurcate in a bifurcation point in $\mathcal{M}$. If we follow the fence $\mathcal{F}^{0+}$ and cross a bifurcation point, then either the minimum or the saddle point that meet at the fence (depending on the type of bifurcation point) changes the bijective area to which it belongs. If we cross the projection of a triple plus bifurcation point in $\mathcal{C}$ from the tropics to the north, then in $\mathcal{A}$ a minimum undergoes a pitchfork bifurcation into two minima and one saddle point. An equivalent bifurcation happens if we cross the projection of a triple zero bifurcation point, in which case the roles of saddle points and minima are reversed.

The mathematical condition for a bifurcation point is as follows. Let $\mathbf{v}_0$ be the eigenvector of the Hessian matrix, cf. equation (5), at the fence with eigenvalue 0. Then, a bifurcation point is a fence point that satisfies

$$(\mathbf{v}_0 \cdot \nabla_\mathcal{A})^3 V = 0. \quad (10)$$

Solving the above equation, we find the projection in $\mathcal{C}$ of a triple plus bifurcation point lying on the fence $\mathcal{F}^{0+}$ at $(\theta, \phi) = (\pi/3, \pi)$. The coordinates in $\mathcal{C}$ of the projection of a triple zero bifurcation point in $\mathcal{F}^{0+}$ are $(\theta, \phi) = (0.381\pi, 7\pi/6)$. The other projections of bifurcation points belonging to the fence $\mathcal{F}^{0+}$ are obtained, for symmetry reasons, via rotations around the $z$ axis by multiples of $\pi/3$. Using the transformation $\theta \to \pi - \theta$ and $\phi \to \phi - \pi$, one finds the projection of the bifurcation points in $\mathcal{F}^{0-}$.

**Bijective areas and the genus of $\mathcal{M}$.** We can thus map each point in $\mathcal{A}$ onto two opposing points in $\mathcal{C}$. The mapping of a point in $\mathcal{A}$ onto the two points in $\mathcal{C}$ will fall either in the north and the south (one point in each region), or both points fall into the tropics of $\mathcal{C}$. This gives the projections $\mathcal{P}_\mathcal{A}$ of the bijective areas of $\mathcal{M}$ into action space. Hence, we can see how the bijective areas are glued together in $\mathcal{A}$ and $\mathcal{M}$. A bijective area is a connected preimage of either the north, the south or the tropics of control space. That is, there is a one-to-one correspondence between the bijective area in $\mathcal{M}$ and its corresponding region in $\mathcal{C}$. Any loop in $\mathcal{M}$ lying entirely in a bijective area is projected onto a zero homotopic loop in $\mathcal{A}$. Hence, in order to achieve intercellular transport, a loop must cross different bijective areas. The neighbouring bijective areas in $\mathcal{M}$ are shown in Supplementary Fig. 3. We can use it to construct the sequence of bijective areas for a given $\mathcal{L}_\mathcal{C}$. For example, consider the loop $\mathcal{L}_\mathcal{C} = (1_\text{N}, 4_\text{N})$. We start in the tropics, where there is only one minimum which is located in $t_+$. Segment $1_\text{N}$ connects $t_+$ to $n_{+2}$, and segment $4_\text{N}$ connects $n_{+2}$ to $t_+$, which closes the loop.

To compute the Euler characteristic of $\mathcal{M}$ and hence its genus, we need to count the vertices, edges and bijective areas, as detailed next. Topologically, the north of $\mathcal{C}$ is a simply connected area (that is, all loops are zero homotopic) with 12 edges (segments of the borders) and 12 vertices (projection of bifurcation points). Each point in the north of $\mathcal{C}$ has 6 preimages on $\mathcal{M}$. Hence, the north of $\mathcal{C}$ contributes with 6 bijective areas, $12 \times 6 = 72$ edges and $12 \times 6 = 72$ vertices to $\mathcal{M}$. A similar contribution comes from the south of $\mathcal{C}$. The tropics of $\mathcal{C}$ is a non-simply connected area, for example, the equator is zero homotopic in $\mathcal{C}$ but not in the tropics of $\mathcal{C}$. To easily compute the Euler characteristic, we need simply connected areas. We make the tropics of $\mathcal{C}$ simply connected by cutting it along a longitude that connects the projection of two bifurcation points, one in each border of $\mathcal{C}$. The total number of edges of the tropics is thus $12 + 12 + 2 = 26$ and the total number of vertices is $12 + 12 + 2 = 26$. There are 4 preimages of the tropics on $\mathcal{M}$. Hence, the total contribution of the tropics of $\mathcal{C}$ to $\mathcal{M}$ is 4 bijective areas, $4 \times 26 = 104$ edges and $4 \times 26 = 104$ vertices.

Next, we glue the bijective areas to form $\mathcal{M}$. Two unglued edges are glued to form a single edge such that the number of glued edges of $\mathcal{M}$ is $(72 + 72 + 104)/2 = 124$. Regarding the vertices, we have $72 + 72 + 104 = 248$ before gluing them in $\mathcal{M}$. We have to subtract 8 vertices that were artificially produced by cutting the tropics of $\mathcal{C}$ in order to have a simply connected area. We have then $248 - 8 = 240$ unglued vertices. There are two types of vertices on $\mathcal{M}$: bifurcation points where 4 bijective areas meet, and pseudo-bifurcation points where 2 bijective areas meet. There are 24 bifurcation points on $\mathcal{M}$. Hence, we need $4 \times 24 = 96$ unglued vertices to glue together the bijective areas at the bifurcation points. The remaining $240 - 96 = 144$ unglued vertices are glued in pairs to form $144/2 = 72$ pseudo bifurcation points. The total number of vertices on $\mathcal{M}$ is the sum of the number of bifurcation and pseudo bifurcation points: $24 + 72 = 96$.

Finally, the Euler characteristic of $\mathcal{M}$ is $\chi(\mathcal{M}) = 96 - 124 + 16 = -12$ and the genus of $\mathcal{M}$ is $g(\mathcal{M}) = 1 - \chi(\mathcal{M})/2 = 7$. Similar arguments can be used to calculate the genus of the submanifolds that form $\mathcal{M}$, that is, $\mathcal{M}_+$, $\mathcal{M}_0$ and $\mathcal{M}_-$. We show in Supplementary Fig. 2 a plaster model of $\mathcal{M}$.

**Computer simulations.** We simulate the motion of point dipoles moving in the potential given by equation (1) using Brownian dynamic simulations. The equation of motion is

$$\xi \dot{\mathbf{x}}_\mathcal{A}(t) = \pm V(\mathbf{x}_A, \mathbf{H}_\text{ext}(t)) + \eta(t), \quad (11)$$

where $t$ is the time, $\mathbf{x}_\mathcal{A}$ is the position of the dipoles in $\mathcal{A}$, $\xi$ is the friction coefficient and $\eta$ is a Gaussian random force with a variance given by the fluctuation-dissipation theorem. The plus (minus) sign in front of the potential holds for the diamagnetic (paramagnetic) colloids. The equation of motion is integrated in time with a standard Euler algorithm. We use a time step $T/dt \approx 2 \cdot 10^5$ with $T$ the period of a modulation loop $\mathcal{L}_\mathcal{C}$. Simulations fully reproduce the experimental phase diagram.

**Data availability.** The data that supports the findings of this study are available from the corresponding author upon request.

### Acknowledgements
We thank Ingrid Bauer, Matthias Schmidt, Pietro Tierno and Antonio Ortiz-Ambriz for illuminating discussions and critical reading of the manuscript. We thank Thomas Hauenstein and Konrad Stern for playing the Franconian Landler with Johannes Loehr. Publication costs have been partially funded by the Profilfeld Polymer- und Kolloid-forschung of the University of Bayreuth.

### Author contributions
J.L. performed the experiments, designed the modulation loops and played the accordion in movie 'Franconian Landler.avi'. M.L. developed the topological relation between $\mathcal{A}$, $\mathcal{C}$ and $\mathcal{M}$. D.de las H. had the idea to use Brownian dynamics to simulate the motion. Simulations were performed by A.E. and led to the discovery of the topological protection of the ratchet. T.M.F. had the idea of the experiments. J.L., D.de las H. and T.M.F. wrote the manuscript.

### Additional information
**Supplementary Information** accompanies this paper at http://www.nature.com/naturecommunications

**Competing financial interests:** The authors declare no competing financial interests.

**Reprints and permission** information is available online at http://npg.nature.com/reprintsandpermissions/

**How to cite this article:** Loehr, J. *et al.* Topological protection of multiparticle dissipative transport. *Nat. Commun.* 7:11745 doi: 10.1038/ncomms11745 (2016).